\documentclass{aa} 
\usepackage{graphicx}
\begin{document}
   \title{Improvements to Existing Transit Detection Algorithms and Their Comparison}

   \author{B. Tingley
          \inst{1}
          }

   \offprints{B. Tingley}

   \institute{$^{1}$Institut for Fysik og Astronomi (IFA), Aarhus Universitet,
              Ny Munkegade, Bygning 520, 8000 Aarhus C\\
              \email{tingley@ifa.au.dk}
             }

   \date{Received X, 2003; accepted Y, 200Z}

   \abstract{
In Tingley (\cite{tingley}), all available transit detection algorithms were
compared in a simple, rigorous test. However, the implementation of the
Box-fitting Least Squares (BLS)
approach of Kov\'{a}cs et al. (\cite{kovacs}) used in that paper was not
ideal for those purposes. This letter revisits the comparison, using a
version of the BLS better suited to the task at hand and made more efficient
via the knowledge gained from the previous work. Multiple variations of the
BLS and the matched filter are tested. Some of the modifications improve
performance to such an extent that the conclusions of the original paper
must be revised.
   }

   \maketitle
%

\section{Introduction}

A determination of the optimal transit detection algorithm (TDA) is 
an essential part of exoplanet searches via occultation, currently
the only technique that can examine large numbers of stars simultaneously
for planets. The use of an improved TDA can have a significant impact on
the returns from an exoplanet search campaign. As an
example of this, the original analysis of the OGLE observations using a
cross-correlation identified 46 transit-like events (Udalski et al.
\cite{udalski1}), while a re-analysis using the BLS of Kov\'{a}cs et al.
(\cite{kovacs}) identified an additional 13 events from the same data
(Udalski et al. \cite{udalski2}). Among the events identified by the
improved TDA was OGLE-TR-56, which is the only one that has been proven
to be a planet unto this point (Konacki et al. \cite{konacki}).
There have been many TDAs proposed in the literature, but there had been no
attempt at a comprehensive comparison until Tingley (\cite{tingley})
(hereafter referred to as Paper I). Therein, different TDAs were compared in
a rigorous, simple test. However, the implementation of the BLS approach
used was not ideal for this task, which resulted in an
artificially poor performance. The algorithm, taken directly from Kov\'{a}cs'
website, did not cover parameter space in the same way as the other
approaches used in that analysis. It had an extra free parameter (the period),
and searching
this extra free parameter greatly increased the false alarm probabilities.
This necessitates a re-visitation of the analysis. Rather than performing
identical simulations to those in Paper I, the information from that paper
could be used to create better simulations that are more demonstrative of
true detector performance and demanding significantly less computational
load. In the process of these
subsequent simulations, it became evident that certain slight modifications
could be made that would improve detector performance.

\section{Comparison method}

The method used to compare the different TDAs here differs from that in Paper
I, taking advantage of the concepts uncovered therein to produce both a
more efficient and more comprehensive test. It was found that the ability of
a TDA to detect a transit depends on the signal energy, $E_S = N_{\rm{in}}
\times
d^2$, where $N_{\rm{in}}$ is the number of observation during transit and $d$
is the depth of the transit compared to the scatter of the light curve.
It should be noticed that this result does not depend on how many different
transits are observed, only the number of observations in transit.
Therefore, it is not necessary to explore parameters such as the number of
transits, the duration of the transits and the period of the transits
as they ultimately have no bearing on detector performance in these
simulations. They act only to inflate the
false alarm probability as a result of the additional tests necessary to search
the expanded parameter space. As long as the number of tests is the same for
the different TDAs, these parameters are irrelevant.

The only parameters that remain to be defined are the transit
signal energy and the total number of observations. The total number of
observations has a relatively small effect that gets smaller as the number
increases and is therefore left fixed for simplicity. This means that
the transit signal energy is the only parameter to be varied. From the
transit signal energy, the number of in-transit observations is chosen
randomly and the depth
calculated. Similar to Paper I, a noise-only light curve is generated
(only white Gaussian noise for these
simulations, unlike Paper I) and copied, with the transit signal added to
one of them. Both are then normalized to have unity standard deviation.
The different detectors are then applied to both light curves, assuming
constant (unity) weights.
This process is repeated $10^5$ times, and from these trials the probability
of detection at a false alarm rate of 1\% for the given signal energy is
determined. This is then repeated
for several different signal energies for completeness.
It should be noted that in this analysis, the transit signal contains only a
single transit, as separating them into multiple transits will only vastly
increase the computational load without improving the comparison.

\section{Transit detection algorithms}

This round of simulations tests fewer transit detection algorithms, as the
comparisons in Paper I are largely sufficient. Both the BLS and the Bayesian
(Defa\"{y} \cite{defay}) suffered from the effects of searching an extra
free parameter, but the BLS outperformed the Bayesian by a wide margin. It is
therefore not necessary to test the Bayesian. Moreover, the matched filter
clearly performed better than the correlation and Deeg's approach (Doyle et
al. \cite{doyle}), so the
latter two can also be left out. Therefore, only the essential approaches
will be included: the matched filter, which was determined
to be the best in Paper I and the BLS, which was not implemented in the ideal
fashion. Small variations on the matched filter and the BLS will also be
included in the comparison.

\subsection{Matched filter approach}

The version of the matched filter used in this analysis is slightly different
from the one in Paper I. The problem with the formulation used there is that
there is one term that is dependent on the number of in-transit observations
that is essentially dropped in the derivation. If one wishes to compare the
likelihood of events that are observed a varying number of times, then this
term is necessary for ideal performance. With this term included, the
equation for the matched filter in the case of a square-well transits is:

\begin{equation}
T = d\sum_{n=\rm{in}}\frac{x_{n}}{\sigma_{n}^{2}} - \frac{1}{2}d^2\sum_{n=\rm{in}}\frac{1}{\sigma_{n}^{2}}
\end{equation}
where $d$ is the depth of the transit, $\sum_{in}$ is the sum over the
in-transit points, $x_{n}$ are the observed differential magnitudes and
$\sigma_{n}$ is the point-to-point noise.

Depth is clearly a free parameter in this formulation of the matched filter.
While tests of depth are not entirely independent, it still requires a fair
number of calculations. The BLS uses a minimization to remove this free
parameter. With this form of the matched filter, it is possible to perform
a similar maximization that will remove this free parameter. $T$ will be
at a maximum when the proper depth is found, so if the equation
is maximized:

\begin{equation}
\frac{dT}{d(d)} = \sum_{n=\rm{in}}\frac{x_{n}}{\sigma_{n}^{2}} - d\sum_{n=\rm{in}}\frac{1}{\sigma_{n}^{2}} = 0.
\end{equation}
Solving for $d$ yields
\begin{equation}
d = \sum_{n=\rm{in}}\frac{x_{n}}{\sigma_{n}^{2}}\left(\sum_{n=\rm{in}}\frac{1}{\sigma_{n}^{2}}\right)^{-1}.
\end{equation}
Substituting this back into the equation for $T$ yields
\begin{equation}
T = \frac{1}{2}\left(\sum_{n=\rm{in}}\frac{x_{n}}{\sigma_{n}^{2}}\right)^2\left(\sum_{n=\rm{in}}\frac{1}{\sigma_{n}^{2}}\right)^{-1}.
\end{equation}
This modified formulation of the matched filter will be referred to as
the maximized matched filter.

\subsection{Box-fitting technique}

As mentioned in Paper I in Section 5, the BLS is based on mathematics very
similar to the matched filter. There are two things that are different about
it. First of all, it accounts for both the in-transit and out-of-transit
levels, rather than assuming that the out-of-transit level is zero as the
matched filter does in the simple, efficient forms presented here. This can be
significant especially for short period exoplanets with deep transits, as the
standard practice of setting the average of the light curve equal to zero can
cause significant deviations in the out-of-transit levels from zero.
However, one should also note that these types of transits are
the easiest to identify. Secondly, it uses a minimization to remove the depth
as a free parameter, which is what inspired the maximized matched filter
shown above. The test statistic for the BLS is
\begin{equation}
T=\frac{s^{2}}{r(1-r)},
\end{equation}
where
\[s=\sum_{n=\rm{in}}w_{n}x_{n},\]
\[r=\sum_{n=\rm{in}}w_{n}, \;\;\; \rm{and} \]
\[w_{n} =\sigma_{n}^{-2}\left[\sum_{m=1}^{N}\sigma_{m}^{-2}\right]^{-1},\] 
where $N$ is the number of observations in the light curve. From inspection,
one can see how closely this resembles the maximized matched filter. The
only significant difference is the $(1-r)$ term, which arises
from accounting for the out-of-transit levels as mentioned above.
Moreover, if one takes the long-period limit (many more
out-of-transit observations than in-transit), $r \ll 1$ and the two equations
become essentially equivalent -- unsurprisingly, since the longer the period,
the closer the out-of-transit level will be to zero.

\begin{table*}
 \centering
 \caption[]{Probabilities of detection for a false alarm rate of 1\% for
various transit signal energies and detectors based on $10^5$ trials. TSE
is the transit signal
energy, MF is the ordinary matched filter, mMF is the maximized matched
filter, mMF$_{\rm{c}}$ is the maximized matched filter with the directional correction,
BLS is the BLS, BLS$_{\rm{c}}$ is the BLS with the directional correction, and uBLS is
the unminimized BLS. The errors quoted are derived by calculating the probability
of detection for a false alarm rate of 1\% for the 10 sets of $10^4$ trials
and determining their standard deviation.}
\small
 \begin{tabular}{c|cccccc}
TSE & MF & mMF & mMF$_{\rm{c}}$ & BLS & BLS$_{\rm{c}}$ & uBLS \\
 \hline
1  & 0.0166$\pm$0.0018 & 0.0134$\pm$0.0006 & 0.0151$\pm$0.0010 & 0.0141$\pm$0.0009 & 0.0157$\pm$0.0014 & 0.0163$\pm$0.0016 \\
2  & 0.0247$\pm$0.0016 & 0.0186$\pm$0.0015 & 0.0232$\pm$0.0011 & 0.0193$\pm$0.0015 & 0.0243$\pm$0.0012 & 0.0242$\pm$0.0018 \\
4  & 0.0526$\pm$0.0039 & 0.0398$\pm$0.0025 & 0.0538$\pm$0.0039 & 0.0420$\pm$0.0030 & 0.0553$\pm$0.0037 & 0.0502$\pm$0.0048 \\
8  & 0.1662$\pm$0.0115 & 0.1305$\pm$0.0061 & 0.1698$\pm$0.0049 & 0.1365$\pm$0.0067 & 0.1767$\pm$0.0061 & 0.1547$\pm$0.0104 \\
16 & 0.5532$\pm$0.0095 & 0.4482$\pm$0.0127 & 0.5257$\pm$0.0112 & 0.4589$\pm$0.0140 & 0.5359$\pm$0.0099 & 0.5522$\pm$0.0100 \\
32 & 0.9363$\pm$0.0048 & 0.9075$\pm$0.0053 & 0.9364$\pm$0.0048 & 0.9136$\pm$0.0044 & 0.9404$\pm$0.0046 & 0.9403$\pm$0.0045 \\
64 & 0.9999$\pm$0.0001 & 0.9997$\pm$0.0002 & 0.9999$\pm$0.0001 & 0.9999$\pm$0.0002 & 0.9998$\pm$0.0002 & 0.9999$\pm$0.0002 \\
 \hline
 \end{tabular}
\normalsize
\end{table*}

Preliminary results showed that the ordinary matched filter outperformed
the maximized matched filter. As the existing formulation of the BLS
includes a similar minimization, a version of
the BLS without the minimization should be investigated. This is
derived quite readily from Eq. 1 of Kov\'{a}cs et al (\cite{kovacs}), realizing
that 
\[\sum_{n=1}^{N} w_{n} = 1 \rightarrow \sum_{n=\rm{out}} w_n = 1 - \sum_{n=\rm{in}} w_n \;\;\; \rm{and}\]
\[\sum_{n=1}^{N} w_n x_n = 0 \rightarrow \sum_{n=\rm{out}} w_n x_n = - \sum_{n=\rm{in}}w_n x_n,\]
where $\sum_{n=\rm{out}}$ is the sum over the points out-of-transit. The
resulting equation for the test statistic, which will be referred to as the
unminimized BLS, is:
\begin{equation}
T= \frac{rd^2 - 2sd}{1-r},
\end{equation}
where $r$, $s$  and $d$ are as above.
As expected, the BLS can be recovered from this by taking $\frac{dT}{d(d)}
= 0$, solving for $d$, and plugging the result back into $T$.

\section{Results}

One interesting result that became apparent during the course of the analysis
was that the maximized matched filter and the BLS both had a slight problem
that hurt their performance. The terms that involved a weighted summation over
the in-transit observations were squared. This removes the sign information
of the summation, meaning that a periodic increase in magnitude will have
the same test statistic as a periodic decrease. The chance of
having random noise mimic a signal of a given energy is thereby doubled,
increasing the overall level of the false alarms.

This can be easily corrected by simply not calculating the test statistics
for any test transit where the weighted sum of the in-transit differential
magnitudes was negative (corresponding to an increase in brightness). This
correction, which will hereafter be referred to as the directional correction,
will not affect the ordinary matched filter, as it depends linearly
on this sum. The
significance of this small change can be seen in Table 1, along with all the
other results. The errors quoted therein are derived by calculating
the probability of detection for a false alarm rate of 1\% for the 10 sets
of $10^4$ trials and determining their standard deviation.

One additional point of concern worth checking is the veracity of the
statement that the transit signal energy as formulated here really
does govern the performance of the detectors. In the simulations used to
create Table 1, the widths and the depths of the inserted transits were
also recorded. The results were binned according to transit width and
no dependence on transit width was observed.

\section{Conclusion}

The uncorrected BLS itself performs considerably better than reported in
Paper I, as expected, but not as well as the ordinary matched filter.
After the directional correction is applied, the BLS demonstrates the best
overall performance of all of the detectors. The maximized matched filter
also experiences significant improvement after the directional correction
is applied, although it does not seem to outperform the matched filter or
the corrected BLS overall. The unminimized BLS performs well, but again
never better than both the matched filter and the corrected BLS
for any of the transit signal energies tested.

It appears from this analysis that the best method of identifying transit-like
features in light curves would be to apply the ordinary matched filter, the
maximized matched filter, the corrected BLS and the unminimized BLS to the
sample of light curves, as no detector is clearly superior for all transit
signal energies.

\begin{acknowledgements}
I would like to thank the Danish Natural Sciences Research Council
for financial support. And finally I would like to thank Hans Kjeldsen
and J\o rgen Christiansen-Dalsgaard for all the support they have provided
me over the past two years.
\end{acknowledgements}


\begin{thebibliography}{}
   \bibitem[2001]{defay} Defa\"{y}, C., Deleuil, M., \& Barge, P. 2001,
     A\&A 365, 330
   \bibitem[2000]{doyle} Doyle, L. R., Deeg, H. J., Kozhevnikov, V. P.,
     et al. 2000, ApJ, 535, 338
   \bibitem[2003]{konacki} Konacki, M., Torres, G., Jha, S., \& Sasselov,
     D.D. 2003, Nature 421, 507
   \bibitem[2002]{kovacs} Kov\'{a}cs, G., Zucker, S., \& Mazeh, T. 2002,
     A\&A 391, 369
   \bibitem[2003]{tingley} Tingley, B. 2003, A\&A 403, 329
   \bibitem[2002]{udalski1} Udalski, A., Paczy\'{n}ski, B., \.{Z}ebru\'{n}, K.,
     et al. 2002, AcA 52, 1
   \bibitem[2002]{udalski2} Udalski, A., Paczy\'{n}ski, B., \.{Z}ebru\'{n}, K.,
     et al. 2002, AcA 52, 115
\end{thebibliography}
\end{document}